\documentclass{elsart}

\usepackage{natbib}

\usepackage{epsfig}

\usepackage{amssymb}

\begin{document}

\begin{frontmatter}



\title{X-ray and UV Variability in the Core of M15: the nature of X2127+119/AC211}


\author[South]{P.A. Charles}
\author[South]{W.I. Clarkson}
\author[Oxford]{L. van Zyl}

\address[South]{Dept of Physics \& Astronomy, University of Southampton}
\address[Oxford]{Dept of Physics, University of Oxford}

\begin{abstract}

The prototypical post-core-collapse globular cluster M15 provides an
ideal environment for the formation of exotic binaries, it being
already known to contain a luminous LMXB (X2127+119, optically
identified with the 17.1 hour eclipsing binary AC211) and six
millisecond pulsars in the core.  However, the X-ray properties of
X2127+119 are strange in that it appears to be a high inclination
accretion disc corona source (ADC) from which we only see scattered
X-radiation, and yet it has produced extremely luminous type I
X-ray bursts, which may or may not have come from AC211.  We have
therefore examined the $\sim$5-year RXTE/ASM light curve of this
object in order to search for long-term modulations which might shed
light on this unusual behaviour, and which led to our discovery of further
X-ray bursts. Furthermore we have used archival HST
UV images of the M15 core to search for other variable objects which
might indicate the presence of a second LMXB. From these we have found
one highly variable ($> 5 mag$) object which we interpret as a dwarf
nova.

\end{abstract}


\end{frontmatter}

\section{Introduction}
\label{}

The dense cores of globular clusters are now well established as
breeding grounds for the creation of exotic interacting binaries
(e.g. Verbunt \& van den Heuvel 1995).  In particular, it has been
noted for almost 20 years that luminous low-mass X-ray binaries
(LMXBs) are found in $\sim$12 globular clusters, an overdensity
relative to the field of a factor 100, indicating a greatly enhanced
formation mechanism in the dense cluster environment (both 2-body
(tidal) and 3-body (exchange) mechanisms are still actively
considered, see e.g. Elson, Hut \& Inagaki 1987 and references
therein).  And the descendants of LMXBs, the millisecond radio pulsars
(MSPs), are also found in substantial numbers in clusters, with 8 of
them in M15 alone (of which 6 are within 8 arcsecs of the core; see
e.g. Lyne 1996).  Furthermore these LMXBs have a range of properties
more extreme than those in the field.  All except one have produced
type I X-ray bursts (e.g. Charles 1989 and references therein), and
hence are accreting neutron stars, but their period distribution is
skewed to include at least two extremely short period binaries
(including the shortest orbital period currently known, X1820-30 in
N6624) which require both members of the binary to be degenerate
(Rappaport et al 1987).  Recent UV surveys with HST (de Marchi \&
Paresce 1994, 1995, 1996) have revealed a population of very blue
stars (VBS) in M15, the nature of which is as yet undetermined, but
their very high density in the core indicates a population segregation
that must be related to other properties of the cluster.

Against this background, M15 houses one of the more enigmatic of these
high luminosity LMXBs, X2127+119 in that it was identified 15 years
ago (Auri\`{e}re et al 1984; Charles et al 1986) with by far the most
luminous optical counterpart (AC211) of any galactic LMXB.  This 15th
magnitude extremely blue, highly variable ($>$1 mag in U) object was
photometrically monitored by Ilovaisky et al (1993) who showed that it
was eclipsing with a 17.11 hour period.  The broad eclipse feature
(seen in both optical and X-rays) suggested that X2127+119/AC211
belonged to the accretion disc corona (ADC) class, in which the high
inclination means that the compact object is obscured by the disc rim
and hence not directly visible.   This accounts for the high
optical luminosity associated with a relatively faint X-ray source.

However, this model suffered a severe blow when Dotani et al (1990)
and van Paradijs et al (1990) observed M15 with {\it Ginga} and
discovered an extremely luminous type I X-ray burst, thereby inferring
that the X-ray source must be directly visible (at least for part of
the time).  In which case the persistent, low apparent X-ray
luminosity ($3{\times}10^{36}$ erg s$^{-1}$ for an assumed distance of
10.5kpc; Djorgovski 1993) was then attributed to a genuinely low
accretion rate ($\sim$0.01L$_{Edd}$), rather than being the result of
obscuration in an ADC, and indeed there are other comparable
luminosity directly-viewed LMXBs in globular clusters (see e.g. Sidoli
et al 2001).  However, these are all optically much fainter and hence
the remarkable optical brightness of AC211 remains unexplained.

We therefore decided to use archival X-ray and UV data on M15 in order
to investigate the strange properties of X2127+11/AC211 in more
detail.  The on-line database of the RXTE ASM provides a unique,
long-term ($\sim$5 yrs), continuous X-ray light curve which we could
examine for behaviour that might be related to the occurrence of the
bursting events (it already having been suggested by Corbet et al 1997
that there might be a 37 day modulation in earlier samples of the ASM
data). Furthermore it was not clear to us that the very compact core
of M15 was restricted to containing only one luminous LMXB, and so we
extracted the HST archival images of the core region in order to
search for variable UV sources, a study that might also shed light on
the nature of the VBS.

\section{RXTE/ASM Light Curve}

Launched in 1996, the Rossi X-Ray Timing Explorer (RXTE) carries an
All Sky Monitor (ASM), which gives continuous coverage of the entire
sky.  Roughly six readings - called ``dwells'' - are taken of each of a
list of sources per day, lasting about 90 seconds per dwell. Timing
information is obtained to within a thousandth of a day, as well as
crude spectral information.  The ASM is sensitive to photon energies
between 1.3 and 12.1 keV, broken into three energy channels
(1.3-3.0~keV, 3.0-5.0~keV and 5.0-12.1~keV).

For weak sources such as X2127+119, which averages only
$\sim$1$cs^{-1}$, particular attention must be paid to problems with
the RXTE environmental background (see figure 1).  To that end we
utilised the software provided by MIT (Levine et al 2000) which sets
levels of background activity for rejection of data which are
significantly better at removing contaminated data than the standard
data product provided on-line. Data was kept only if the fit had a
$\chi^2$ value between 0.05 and 1.5, and the background at the time of
the dwell was less than 10 $cs^{-1}$. This produces a substantial
improvement in the data quality.

To identify longterm trends in the ASM lightcurve of a weak source
such as X2127+119, it is instructive to bin the data, and we chose a
binning of five days per bin and required a minimum of five datapoints
ber bin. The resulting lightcurve and rms error are shown in figure
2. We see regions of extended activity at about day number 160 and
460, lasting $\sim$50d in each case.

\begin{figure}
\begin{center}
\psfig{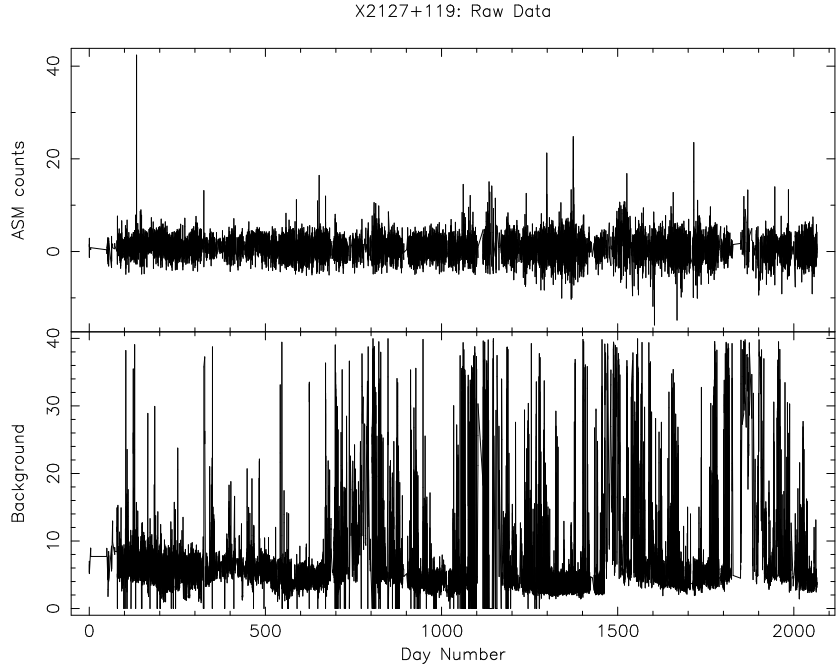}
\end{center}
\caption{RXTE ASM lightcurve of X2127+119 and background before cleaning. Day numbers are with respect to Jan 6, 1996.}
\label{fig1a}
\end{figure}

\begin{figure}
\begin{center}
\psfig{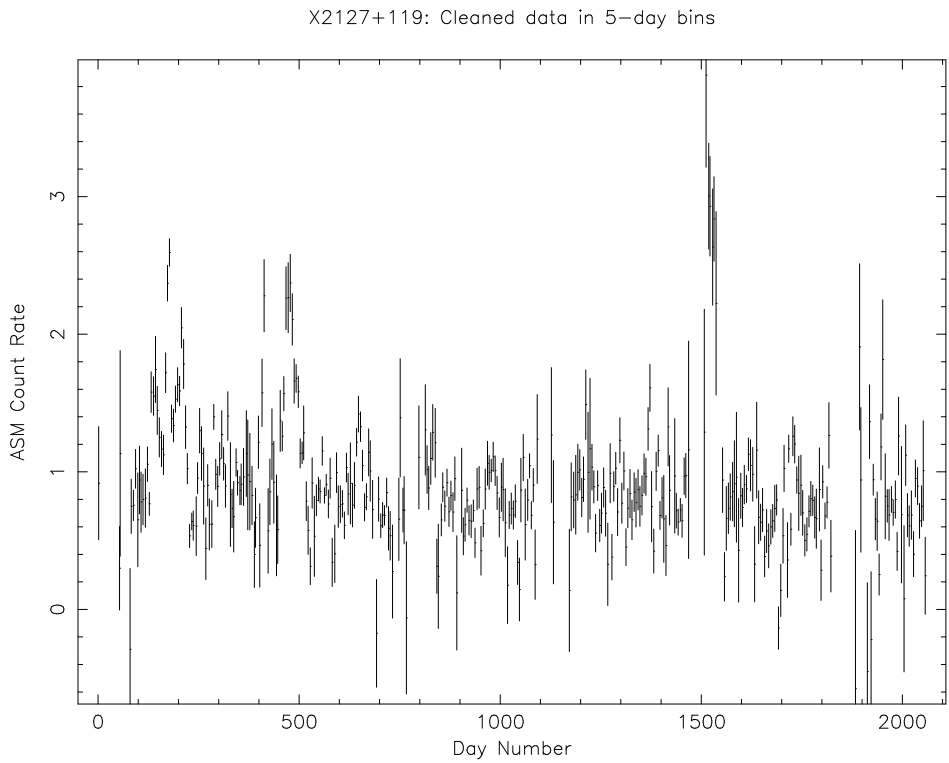}
\end{center}
\caption{The RXTE ASM dataset, but with all points with background levels above 10 counts per second removed, all points with $\chi^2>$1.5 removed. Data has been binned to 5 days per bin to highlight any intervals of extended ray activity in X2127+119.
}
\label{fig2}
\end{figure}

\subsection{Period Search}

Because the ASM data is unevenly sampled, it was necessary to replace
the Fourier Transform with the Lomb-Scargle periodogram (LS) when
searching for periodicities (Scargle, 1982). The LS periodogram is a
Discrete Fourier Transform modified to take account of unevenly
sampled data. Phase Dispersion Minimisation (PDM) detects
periodicities by folding the data on trial periods, producing a
measure of the scatter in the resulting fold. Because any periods
found by this method do not depend on the form of the underlying
modulation, PDM is a useful check on any periodicities found with LS
techniques and is used here to provide an independent measure of the
period. The resulting periodograms are shown in Figures 3 and 4.

With a source as weak as X2127+119 it is necessary to determine the
significance of detected periods. This is accomplished by performing
the period analysis on a large number of simulated datasets, where the
simulations are composed of Gaussian noise on the sampling of the ASM
data. The Lomb-Scargle power in the resulting power spectrum of each
dataset is calculated, allowing us to measure the correspondence
between the power of an LS peak and the probability that the peak
arose from a genuinely periodic signal, and was not due to noise. From
our simulations, we find that an LS power of $\sim$12 corresponds to
99 percent confidence that the power is due to a real signal, assuming
white noise (for full details see Homer et al, 1996).

Figure 3 shows the period analysis performed over long periods, from 2
to 1000 days. For the purposes of period detection, the data was
binned up to one day averages. There is no sign from this analysis of
the 37d period reported by Corbet et al 1997. By adding sinusoidal
signals to our simulated light curves, a rough upper limit was set for
the amplitude of a 37d periodic signal that might go undetected with
the time sampling of the ASM dataset. This upper limit occurs at about
0.08 $cs^{-1}$, or roughly 8 percent of the average quiescent level.

It must be borne in mind that this analysis was carried out for the
full energy range of the ASM; whereas Corbet et al point out that this
modulation is only prominent in ASM Channel 1. Therefore
power spectra were computed for the individual channels
(figure 4), and these also showed no significant periodicities. A plot of the
log-log binned power spectrum shows that white noise is a good model
for individual channel data. The 37d period can therefore be safely
dismissed as an artefact.

Recent analysis of ASM data for the dipper X1916-053 (Homer et al.,
2001), however, shows that significance levels based on a white noise model may
not be valid for the total ASM count rate, and the noise can be frequency
dependent. Plotting the binned power spectrum on a log-log scale shows
that the noise is better described by the sum of a power law ``red
noise'' component and a white noise component, with a break at
frequencies $\sim$ 0.02 cycles per day. The result is that the LS
power corresponding to a 99$\%$ significant periodicity detection
varies with frequency. With this in mind, model spectra were computed
using the best-fit red-noise model for X2127+119, following the method
of Uttley et al (2001).  The structure in the power spectrum was found
to be entirely consistent with this noise model. Consequently, we find
no candidate periodicities distinguishable from noise.

\begin{figure}
\begin{center}
\epsfig{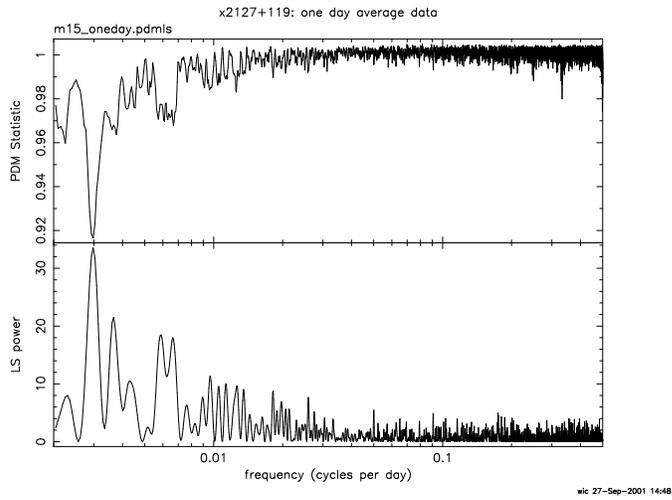}
\end{center}
\caption{Periodograms over a range of periods from 2 to 1000 days. There is no evidence of the previously suggested 37d period (see text).}
\label{fig3}
\end{figure}

\begin{figure}
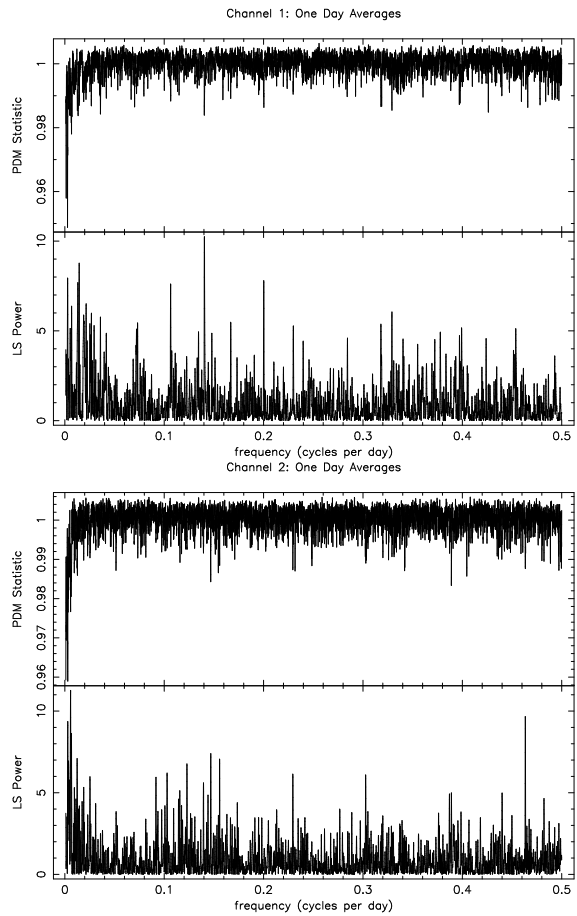

\begin{center}
{\psfig{file=per_ch1.ps,width=6cm, angle=-90}
\psfig{file=per_ch2.ps,width=6cm,angle=-90}}
\end{center}
\caption{Period analysis of X2127+119 performed on the lower two channels.}
\label{fig4}
\end{figure}

Finally, a similar period analysis performed over a range of slightly
shorter periods confirms that the 17.1 hour orbital period of AC211
continues to appear in the ASM dataset. This has been analysed in more
detail elsewhere; Homer \& Charles 1998 examined the first 2 years of
ASM data in search of a period derivative, and more recently Ioannou
et al. (2001) performed a similar search on the full five year
dataset.

\subsection{X-ray burst candidates}

The coverage of the ASM is such that a few individual dwells (of 90s
each) are obtained each day, at an average rate of about one every few
hours. This sampling rate is an order of magnitude longer than typical
type I X-ray burst durations - the October 1988 burst, for example,
lasted less than 200s. However, given the extent of the ASM database,
we might still expect to see a burst event in the ASM lightcurve, as a
single, high datapoint. With this in mind, a search was undertaken for
similar X-ray bursts to the October 1988 event.

While selecting for background, the cleaning procedures detailed in
the previous section also select the data based on minimising a least
squares fit. While appropriate for most analyses of RXTE data, this
tends to reject single high points - the target of this search for
bursts! We therefore decided to refilter the raw data based on
behaviour of the background levels; all points with a background level
greater than 20 counts per second were rejected. A burst is then
recognised if the burst profile differs from the background profile,
whatever the absolute value of the background. All points with no 
background information were rejected.

At $\sim1$$cs^{-1}$, the quiescent output of X2127+119 is so low that with
the exception of burst candidates, which have $\sim$5 times the
signal-to-noise ratio, variations in the shape of the lightcurve are
indistinguishable from noise. This is especially true of data for the
individual channels, with values typically $\sim$1/3 $cs^{-1}$. For this
reason, the shape of the lightcurve either side of a burst candidate
is useless when attempting to discern bursts; another measure must be
found that uses only the burst candidate events themselves. The method
chosen to examine candidates was to compare their approximate peak
photon energies.

The classical type I X-ray burst profile consists of a fast rise to
peak luminostity, (rise time typically $\sim$1 - 10s), followed by a
more gradual decay to quiescence (decay time typically from $\sim$ 10s
to $\sim$1 minute). Decay times are generally much shorter at higher
photon energies, producing a softening of the burst profile during
decay that is ascribed to the cooling of the neutron star photosphere
after the initial detonation (Lewin et al, 1995).

The implication for any burst search is that we should expect to see a
contribution to the burst from all three ASM channels. This allows us
to reject burst candidates with activity in one channel only. A
further implication is that we should see a rough anticorrelation
between total luminosity and hardness ratio as defined in this paper
(section 2.5), as the more luminous points would arise from earlier
points in the profile with higher photon energies.

\subsection{The clockwork burster X1826-238: a comparison source}

To demonstrate the validity of our search strategy, the X-ray burster
X1826-238 was examined with the ASM. X1826-238 displays a remarkable
near-periodicity in its burst behaviour, with seventy bursts observed
by the BeppoSAX Wide Field Camera over 2.5 years. Analysis by Ubertini
et al (1999) revealed a significant quasi-periodicity in the bursts of
5.76hr (with a FWHM to the distribution of only 0.617hr).

The periodicity of this source allows us to predict possible windows
in which we might expect to see a burst, given the precise timing and
periodicity from Ubertini et al. A burst candidate for X1826-238 is
plotted from the ASM lightcurve in Figure 5: as can be seen, it fits
within the expected window very well. The energy distribution is
weighted towards the higher energies, which supports the standard
model for type I bursts given that the candidate occurs early in the
predicted burst window.

If the method of searching for a rough luminosity-hardness ratio
anticorrelation is valid, then X1826-238, with its dependable burst
activity, should be the ideal calibration source. Plotting the ASM
count rate against hardness ratio for this source does indeed show
such a relation; the points with the highest count rate correspond to
those with the hardest photon energies.

At $\sim$ 2 ASM $cs^{-1}$, the quiescent level of X1826-238 is similar to
X2127+119, making it a good comparison source. Figure 6 shows the
count rate - hardness ratio plot for both X1826-238 and our X2127+119
burst candidates.

\subsection{Previous bursts and their implications for the ASM search}

We must also consider profiles that are a little more varied than the
classical type I profiles exhibited by X1826-238, as the October 1988
burst (and the recent event reported by Smale 2001) was by no means a
typical type I burst. To begin with, after ten years, this burst is
still one of the most luminous X-ray bursts seen, which at
$\sim$4.5$\times$10$^{38}$ erg~s$^{-1}$ is near-Eddington for this
system. (By considering gravitational redshifting at the neutron star
surface, van Paradijs et al (1990) obtain a figure of
$\sim$5$\times$10$^{38}$ erg~s$^{-1}$ for the Eddington limit of this
system). At $\sim$ 150s, the burst was unusually long for a classical
type I burst, and there was also a precursor burst $\sim$ 6s before
the main burst. Finally, the profile was unusual for a type I burst,
with the peak occurring later, at successively higher energies,
causing a {\it hardening} of the spectrum. These properties are
thought to be hallmarks of photospheric radius expansion on the
neutron star as a result of the burst (Lewin et al., 1995)

A second burst was detected with the PCA in September 2000 (Smale,
2001), that shares many characteristics with the October 1988
event. Like the 1988 burst, the profile is double-peaked, indicating
photospheric expansion. Though the overall timescale for both bursts
is similar, the 2000 burst displays shorter timescales for the
photospheric expansion events, suggesting a smaller photospheric
radius at maximum expansion. Unfortunately, due to ASM sampling (see
Section 2.6), this burst was not observed by the ASM.

\subsection{The X2127+119 burst candidates}

Table 1 lists the candidate bursts that satisfy our search
criteria. An event was selected as a burst if its ASM count rate was
above 10 $cs^{-1}$ and the hardness ratio between 0.5 and 5. In each
case the bursts take place in the absence of any significant
background events, and are detected in each energy channel.

To compare candidate burst fluxes with previous bursts, it is
necessary to relate PCA count rates to ASM count rates. This was
accomplished through examination of publicly available ASM data and
archival PCA observations of the steady output of the known burster
X1735-44. The steady output was chosen rather than its bursts for the
reason that no bursts were found in any of fourteen known bursters for
which the same burst is detected in both the PCA and the ASM, whereas
one can be confident that the steady output measured by the PCA is due
to the same source behaviour as the steady output recorded by the ASM
at a different point on the same day. Furthermore, the spectral
properties of these sources are similar. Based on the steady output
from X1735-44, an ASM count rate of 1 $cs^{-1}$ was found to
correspond to an average PCA count rate over 90s of 50.5 $\pm$ 2.9
$cs^{-1}$. We estimate the total PCA counts recorded during the
September 2000 burst at 370000 $\pm 5\%$, which would then correspond
to an ASM count rate of 81 $\pm$ 6.2 $cs^{-1}$. This does assume that
the 90s dwell overlaps perfectly the $\sim$80s burst profile (which as
we shall see in Section 2.6, is a large assumption).

The flux values given in the table are fractions of the peak flux of
the September 2000 burst ($\sim$6.5$\times$10$^{38}$
erg~s$^{-1}$). These values depend on both the source behaviour and
the amount of overlap between dwell and burst, so cannot be taken as
indicators of either quantity on its own. The hardness ratio used here
is defined as the ratio of the count rate in channels 1 and 2 to that
in channel 3 (strictly speaking, a ``softness ratio''). The maximum
duration of the burst was estimated from the data sampling each side
of the event.

Galactic bulge sources have been discovered recently to exhibit
extended X-ray bursts lasting several hours (e.g. Cornelisse et al
2000). The fact that we can only set rough limits on our burst
candidate durations allows the possibility that an event may last
longer than the few minutes of the October 1988 burst; if neutron star
photospheric radius expansion is a recurrent phenomenon this might be
expected. With the time resolution afforded us by the RXTE ASM, we can
do little more than admit the possibility that some of our candidates
may be consistent with long bursts.

The fluxes hover around 10 to 20\% of the October 1988 burst, and
there is quite a wide range of hardness ratios and durations for the
candidates. The coverage of candidate event 11 is such that there are
actually three points recorded that may be part of the same profile;
this candidate is plotted alongside a probable burst from X1826-238 in
figure 5.

Figure 6 shows the count rate versus hardness ratio plots for both
sources. The qualitative similarity gives us some confidence that we
are indeed seeing burst profiles sampled at random points from
X2127+119; all points for which the low energy channels dominate also
show low count rates. This is further strengthened by the behaviour in
figure 6 of the variations in the known non-burster X1700-37, which
show very little spectral changes.

\subsection{ASM sampling and the true burst rate}

Each dwell measurement lasts 90s, and there are of order 10 dwells per
day. To be detected, a significant fraction of the $\sim$80s burst
profile must overlap with a 90s dwell, so that the average count rate
of the 90s dwell is high enough to stand out from noise.  Making the
assumption that the burst profile of September 2000 (Smale, 2001) is
typical, the probability of burst detection can be calculated and
related back to the number of bursts measured, so as to make an
estimate of the true burst recurrence time.

Setting the lower limit for a burst to stand out from noise at 10
$cs^{-1}$, we find that the average count rate over a 90s dwell must be
at least 12$\%$ of the peak count rate if detection is to occur. This
count rate is reached if either (i) the end of the dwell period occurs
4s or more into the burst profile, or (ii) the beginning of the dwell
occurs 49s or more before the end of the burst profile.

Using the known average interval between dwell measurements and
treating the start and end times of burst and dwell as random
variables, we find that the probability of detection is $\sim$ 0.0234
divided by the length of time between bursts. Given that 14 burst
candidates were detected in the $\sim$5 year dataset, this places a
lower limit on the burst recurrence time of $\sim$1.9 days.

\small
\begin{table}
\begin{center}
\caption{RXTE ASM Burst candidates}

\begin{tabular}{lccccl}
\hline
{\em Event} & Day $^1$ & Flux$^2$ & hardness ratio $^3$ & Max. burst duration (mins) \\ 
\hline


1 &     587.946632 &  0.14 & 0.5	&  749.6 \\
2 &  647.897373 &  0.14 &  1.4	&  96.5 \\
3 &  652.492373 &  0.20 & 0.85	&  562.9\\
4 &  1240.353114 &  0.16 &  1.7 	&  96.0 \\
5 &  1338.510521 &  0.14 & 1.0 	&  430.0\\
6 &  1367.671632 &  0.16 &  1.9	&  92.3\\
7 &  1373.595335 &  0.31 & 0.6 	&  93.9 \\
8 &  1374.19441 &  0.22 & 0.6	&  89.6 \\
9 &  1375.124595 &  0.17 &  4.2	&  95.7 \\
10 &  1375.255521 &  0.12 &  2.3	&  4.8 \\
11 &  1715.607188 &  0.30 &  1.1	&  34.9 \\
12 &  1726.483298 &  0.14 &  2.1	&  93.6\\
13 &  1857.959595 &  0.15 &  3.2	&  860.0 \\
14 &  1945.611071 &  0.17 &  1.3	&  95.7 	\\
15 &  1984.320503 &  0.16 &  1.3	&  96.3  \\

\hline

\end{tabular}

\end{center}
{\footnotesize $^1$ Days after January 6th, 1996} 

{\footnotesize $^2$ In units of the peak flux of the September 2000 burst} 

{\footnotesize $^3$ Defined as (1.3-5~keV/5-12.1~keV)}

\end{table}
\normalsize


\begin{figure}
\begin{center}
{\psfig{file=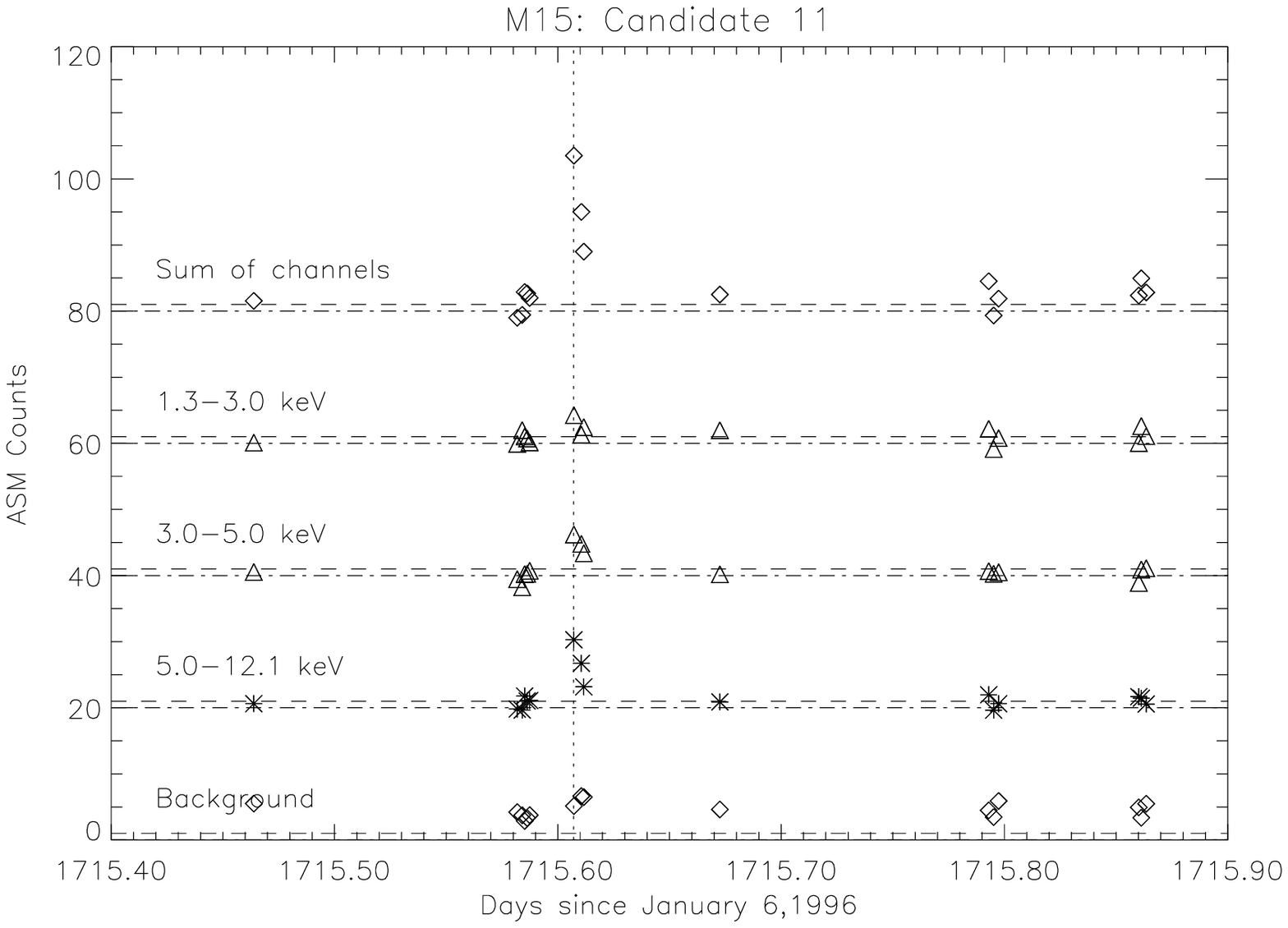,width=6cm}
\psfig{file=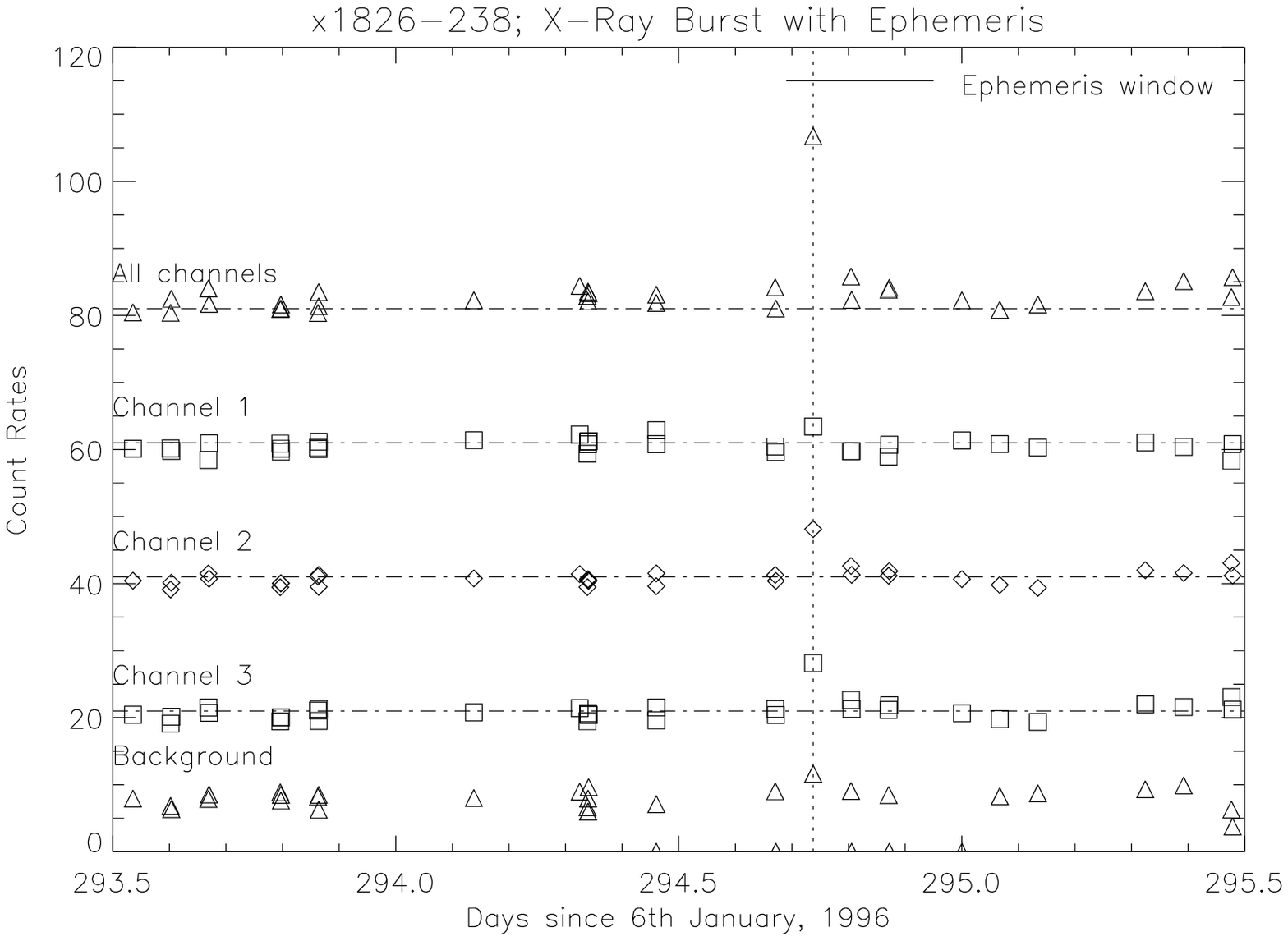,width=6cm}}
\end{center}
\caption{
Burst candidates from X2127+119 (left) and the comparison source X1826-238 (right). The top plot represents the total ASM count rate, beneath this are the three ASM channels individually, in increasing order of hardness. At the bottom is the background behaviour. All plots are to the same scale. The broken lines on the plots represent 0 and 1 for each plot. The solid line above the X1826-238 candidate represents the predicted window for occurrence of a burst, based on the 5.76hr SAX ephemeris. Day numbers are with respect to Jan 6, 1996. 
}
\label{fig5}
\end{figure}

\begin{figure}
\begin{center}
\psfig{file=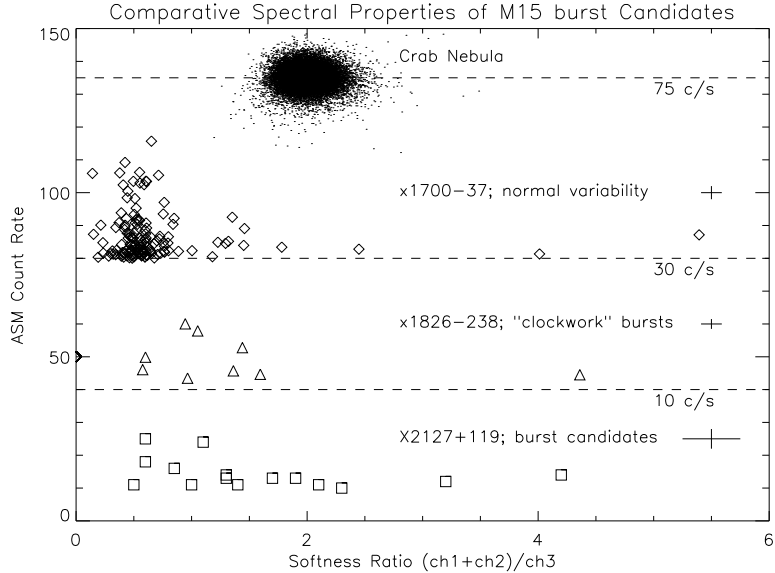,width=11cm}
\end{center}
\caption{Peak ASM count rate plotted against hardness ratio, for the Crab Nebula (top, offset by 65 $cs^{-1}$), brightest points from the known {\it non}-burster X1700-37 (offset by 50 $cs^{-1}$), the burst candidates from X1826-238 (offset by $30 cs^{-1}$) and burst candidates from X2127+119 itself (bottom, no offset). We find that this analysis does indeed differentiate bursting sources from simple variabilities, strengthening our confidence that the candidates selected for X2127+119 do indeed represent snapshots of type I-like X-ray bursts.
}
\label{fig6}
\end{figure}

\section{HST UV Imaging: a variability study}

With its extremely bright and crowded central regions, only the
brightest members of the core of M15 have been studied with
ground-based telescopes (even then requiring the best natural seeing
conditions e.g. Auri\`{e}re \& Cordoni 1981).  Consequently M15 has been a
regular target for the pre- and post-COSTAR corrected HST.

The most detailed optical/UV HST studies of the central stellar
populations of M15 are presented by De Marchi \& Paresce (1994; 1996)
and Guhathakurta et al (1996) (hereafter DMP94, DMP96 and G96
respectively).  DMP94 used far-UV and U band FOC images to derive
colour-magnitude diagrams (CMDs) and demonstrate that there are a significant
number (15) of very blue stars (hereafter VBS) in the M15 core.  DMP96 show
that these are located beyond the blue-straggler section of the CMD,
and have properties similar to sdB or extreme horizontal branch stars
(based mainly on colour and luminosity), but note that detailed
spectroscopic study is necessary in order to elucidate their true
nature.

However, the VBS have properties similar to active LMXBs (indeed one
of them is actually AC211 itself), and the fainter ones could be
accreting white dwarf binaries (i.e. CVs).  Since these interacting
binaries are known to be highly variable, we decided to de-archive
these images and select for further study those obtained of the same
regions with the same instrumental configurations, but separated in
time by long intervals.  In order to find blue variables in the core
of M15, we analysed two sets of HST WFPC2 narrow-U archival
images. The two sets of observations were made in April 1994
(UBVRI observations) and in October 1994 (U-band only).

\begin{figure}
{\psfig{file=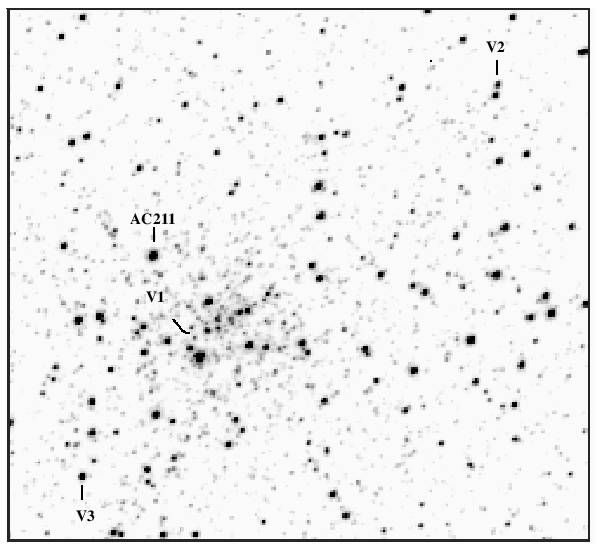,width=7cm}\psfig{file=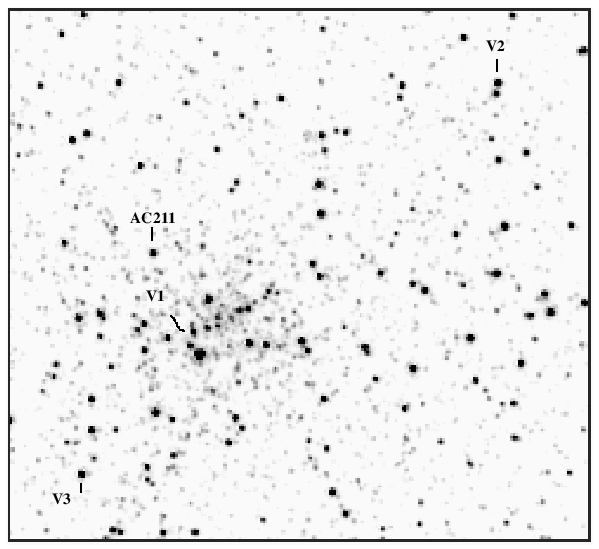,width=7cm}}
\caption{HST WFPC2 U-band images of the core of M15, obtained in April
1994 (left) and October 1994 (right). The positions of AC211 and the
three new variables are indicated. Each image is 11.5 arcsec across;
North points to the top left hand corners, East to the bottom left corners.}
\label{fig10}
\end{figure}

After careful alignment of the images we used {\it iraf} to form
difference images in order to highlight any variable objects.  We
detected four blue variables in an 8-arcsec radius from the central
cusp (see Table 1). One is the known LMXB AC211, but none of the other
three (which we have called V1 - V3) is a previously noted VBS (which
can be most clearly seen in the G96 composite colour image). We used
the Vegamag photometric zeropoints for the WFPC2 to determine the
magnitudes listed below. The limiting U-magnitude in these exposures
(400 s in the F336W filter) is about 22.5.  The variables appeared in
each of the 8 U-band exposures taken in October 1994 and are therefore
not cosmic rays or chip blemishes.  These 3 sources (which varied by
$>$0.8 mag between the observation epochs), are marked in Figure 7. We
have not corrected for interstellar reddening, which in the direction
of M15 is a very small effect; E(B-V) for M15 is 0.09 (Rosenberg et
al. 2000).

\begin{table}
\begin{center}
\caption{HST/UV Variables in the Core of M15}
\begin{tabular}{lccccl}
\hline
 & \multicolumn{3}{c}{Apr 1994} & Oct 1994 & \\ 
{\em Star} & U & B & V & U & {\em Location}$^*$\\ \hline

V1 & $>$22.5 &    - &   - &      17.60 $\pm$ .03 & $\sim$0.3'' NE \\
V2 & 16.49 $\pm$ .01 & 16.20 $\pm $.01 &  15.86 $\pm$ .02 &  15.70 $\pm$ .01 & $\sim$3.5'' due E \\
V3 & 17.20 $\pm$ .02 & 16.97 $\pm$ .02 &  16.56 $\pm$ .04 & 16.07 $\pm$ .02 & $\sim$7.5'' due W \\
  
\hline

\end{tabular}
\end{center}
{\footnotesize $^*$ with respect to cluster centre}
\end{table}

On careful examination, it became clear that the variability
of one of these sources (V1 in Figure 7) was due to the appearance of
a transient event in the U-band images taken in October 1994.  There is no
trace of it in the April 1994 images and so the variability range is
actually $>$5 mags. There is also no sign of it in any other HST
images of M15 (August 1994, December 1998 or August 1999). V2 and V3
on the other hand, are likely to be RR Lyr variables given their
variability range, colours and absolute magnitudes (Allen 1973).

\subsection{The nature of the transient V1}

Although we have no colour information about V1, we expect it to be a
UV-bright object: red variables are likely to be evolved stars and
much brighter than V1 in quiescence. Also, faint stars that exhibit
5-magnitude outbursts are likely to be blue objects.
UV-bright transients in the cores of globular clusters have a strong
possibility of being interacting binary systems containing compact
objects, i.e. cataclysmic variable stars (in which the accreting
object is a white dwarf) or X-ray binaries (neutron-star
systems). Both these classes of variable stars have subclasses which
exhibit transient behaviour. Dwarf Novae (DNe) are faint CVs (M$_{\rm V}$
typically 6 to 8 in quiescence, with a large scatter; see Warner 1995)
which undergo periodic outbursts of 3 to 8 magnitudes. X-ray
transients are X-ray binaries which have similar quiescent absolute
magnitudes to DNe, and which undergo infrequent outbursts of 6 to 9
magnitudes (see e.g. Charles 2001 and references therein).

DNe and X-ray transients in M15 (distance 10.2 kpc, distance modulus
15.31; Rosenberg et al. 2000) would be expected to have quiescent
magnitudes of 21 to 24, with outburst magnitudes ranging from 14 to 21
for DNe, and from 13 to 18 for X-ray transients. At 17.6 mag, the
transient V1 is well within the expected range of peak magnitudes for
both DNe and X-ray transients, which makes it an event long sought after
but rarely seen in the high density environments of globular cluster
cores.  However, we have no idea of knowing whether we have observed V1 at
its peak magnitude, nor was any X-ray ASM operating at this time (RXTE
was launched in Dec 1995) to give us any X-ray coverage.

If V1 is indeed a DN or an X-ray transient, it is statistically much more
likely to be the former than the latter. DNe outburst recurrence times
are weeks to months, while X-ray transients recur on much longer
timescales, typically tens of years. The chances of catching an X-ray
transient in outburst on the few occasions on which the HST has
observed M15 over the last decade are extremely small. 

The probability that V1 is a DN rather than an X-ray transient depends
also on the relative numbers of these two kinds of object in the core
of M15. While X-ray binaries are overabundant by a factor of $\sim$
100 in globular clusters compared to the galactic plane, CVs appear to
be {\em under}abundant by a factor of $\sim$ 1000. Whether this
underabundance indicates that globular clusters are unfavourable
environments for the formation and/or continued existence of CVs, or
whether we simply have not been able to find them in the extremely
challenging observing conditions of globular cluster cores, is a
subject of intense debate (Livio 1996).

There is already an X-ray binary in the core of M15 (AC211) and while
these objects are rare there is no reason for there not to be more
than one in M15. The core of M15 contains several millisecond pulsars
(Phinney 1996), end-products of LMXB evolution, and so we
cannot rule out that V1 is an X-ray binary, even though the chances of
catching a transient in outburst in a single HST visit are extremely
small. What are the chances of detecting a DN in outburst in M15
during a single HST visit? Di Stefano \& Rappaport (1994) predict a
population of $\sim$ 200 CVs in 47 Tuc, of which $\sim$ 50 might be DNe. We
would expect M15 to have a higher population of CVs than 47 Tuc because of
its greater core mass and central density, which would lead to a
higher formation rate of close binaries through tidal capture and
three-body interactions. If we apply Di Stefano \& Rappaport's
predictions to M15, we can assume that it has a population 50 DN. If
these DN have a typical mean outburst recurrence time of 50 days and
outburst length of 10 days, then in any given HST visit, one would expect
to see $\sim$10 DNe in outburst. Di Stefano \& Rappaport's
predictions may be over-optimistic, but they do indicate that the
chances of finding a DN in outburst should be very good, and we
conclude, therefore, that V1 is most likely to be a DN.

\section{Discussion}

The optical and X-ray light curves of X2127+119/AC211 (Homer \&
Charles 1998; Ioannou et al 2001) contain the very stable (and
extended) eclipse feature which requires a high inclination
($\geq$70$^\circ$) for this binary system.  With its very low
L$_X$/L$_{opt}$ ratio ($\sim$20) this has provided very strong support
for the ADC interpretation.  But the van Paradijs et al (1990) X-ray
burst from M15 is the major key factor that does not fit into the ADC
model and overall scenario.  Note also that the distance to M15 is
well constrained at 10.5~kpc, thereby providing accurate luminosities
for both the quiescent and burst X-ray emission.  We consider then
that there are only 3 possible explanations available to resolve this
situation:

(a) the X-ray burst is actually scattered in the ADC but remains
detectable as a coherent event.  We believe that this is unlikely
because the time profile is typical for a type I X-ray burst, and the
peak luminosity is extremely high (at L$_X\sim$4.5$\times$10$^{38}$
erg~s$^{-1}$ assuming isotropic emission; van Paradijs et al 1990) in
spite of the considerable X-ray scattering and reprocessing that would
otherwise be occurring (and restricts ADC sources to only a few
percent of their intrinsic luminosities);

(b) there exists a long-term modulation in the X-ray light curve that
could arise from a warped accretion disc (as a result of an
instability to X-ray irradiation; see Wijers \& Pringle 1999 and
Ogilvie \& Dubus 2001), allowing an occasional direct view of the
inner disc/compact object.  We believe that this is also unlikely for
several reasons.  If the X-ray source were to become directly visible
at certain times, then, assuming an average L$_X$/L$_{opt}$ ratio for
LMXBs (van Paradijs \& McClintock 1995), the increase in X-ray
brightness would be much more than the factor 2 or 3 visible in the
ASM lightcurves (figure 1).  Furthermore, the source would need to be
bursting very frequently in order for an event to occur during a
(rare) period of direct viewing.  And there is simply no evidence for
such frequent bursting in the extensive ASCA and RXTE monitoring
campaigns (Ioannou et al 2001), which are consistent with our burst
frequency limit calculated in section 2.6;

(c) there exists another LMXB (the burster) within the core region of
M15 and also very close to AC211 (within 5 arcsecs).  This is both
expected (on the basis of the compact, high density nature of the core
combined with the effects of mass segregation) and required (in order
for the source not to have been resolved by previous imaging X-ray
missions, particularly ROSAT).  Of course, the X-ray burster could
also be a chance alignment with the direction of M15, but this is
extremely unlikely because of the high galactic latitude of M15
($b_{II}\sim$-27$^\circ$) and the necessarily very precise alignment
required (arcseconds) with the core of M15.  See also the similar
suggestions made by Grindlay (1992, 1993).

We consider therefore that a second active LMXB in M15 is the most
likely explanation and also suggest that such a source might be
associated with one of the UV variable stars we have found in the
archival HST images.  For the observed X-ray flux levels (now
indicative of the intrinsic luminosity of the source) we may scale
with respect to similar LMXB bursters in the plane (e.g. X1735-444,
which has M$_V$ of 2.2, but is ten times more luminous in X-rays; see
van Paradijs \& McClintock 1995) to infer V$\sim$19.  This additional
source could then be responsible for the variations present in the ASM
light curve (figure 1), and hence is likely to be comparable to the
currently observed flux from X2127+119, but of course must be much
fainter than the intrinsic luminosity of the ADC source AC211.

Note that the observed X-ray, optical and UV 17.1hr modulation of
X2127+119/AC211 demonstrates that a significant fraction of the total
M15 X-ray flux must originate from this system, but if it is
contaminated by another source then this implies that the X2127+119
eclipse must be deeper.  There would also be implications for the
spectral properties of the source (Ioannou et al 2001).  To test this
interpretation will require much higher spatial resolution X-ray
imaging of the core of M15.  Such imaging has been approved to be
undertaken later this year with the Chandra HRC (which has a spatial
resolution of $\sim$0.5 arcsecs), the results from which will be the
subject of a future paper.$^*$

{\footnotesize
{\bf{Acknowledgements}}

PAC would like to thank Marty Weisskopf for stimulating discussions
about the nature of AC211 which led directly to some of the work
reported in this paper, Lee Homer for comments and the use of his ASM
analysis software for the period search, and Nick White for informing
us of his Chandra results in advance of publication.  We also thank
Phil Uttley for assistance with the rednoise modelling. LvZ
acknowledges the support of scholarships from the Vatican Observatory,
the National Research Foundation (South Africa), and the Overseas
Research Studentship Scheme (UK). This research has made use of data
obtained through the HEASARC online service, provided by the
NASA/Goddard Space Flight Center.  In particular, we thank the
ASM/RXTE teams at MIT and at the RXTE SOF and GOF at NASA's GSFC for
provision of the ASM data.  WIC acknowledges the support of a PPARC
Research Studentship.  Based on observations made with the NASA/ESA
Hubble Space Telescope, obtained from the data archive at the Space
Telescope Science Institute.  STScI is operated by the Association of
Universities for Research in Astronomy, Inc. under NASA contract NAS
5-26555.

$^*$ In completing this paper we became aware of a
Chandra observation by White \& Angelini (2001) that had
serendipitously resolved X2127+119 into 2 separate sources, confirming
the suggestion made here. We note, though, that the second source is
{\em not} V1, strengthening our interpretation of V1 as a dwarf nova.}


\end{document}